\documentclass{article}
\usepackage{arxiv}

\usepackage[utf8]{inputenc} 
\usepackage[T1]{fontenc}    
\usepackage{hyperref}       
\usepackage{url}            
\usepackage{booktabs}       
\usepackage{amsfonts}       
\usepackage{nicefrac}       
\usepackage{microtype}      
\usepackage{lipsum}
\usepackage{float}
\usepackage{graphicx}
\usepackage{multirow}
\usepackage{array}
\usepackage{comment}
\usepackage{longtable}
\usepackage{ragged2e}
\usepackage{subfig}
\usepackage[table]{xcolor}  
\usepackage{geometry}
\usepackage{multicol}
\usepackage{amsmath}
\usepackage{pifont}
\title{Multimodal Bearing Fault Classification Under Variable Conditions: A 1D CNN with Transfer Learning}

 \author{
    Tasfiq E. Alam \\
    Department of Industrial and Systems Engineering \\
    University of Oklahoma \\
    Norman, Oklahoma-73071 \\
    \texttt{tasfiq@ou.edu} \\
    \And
    Md Manjurul Ahsan \\
    Department of Industrial and Systems Engineering \\
    University of Oklahoma \\
    Norman, Oklahoma-73071 \\
    \texttt{ahsan@ou.edu} \\
    \And
    Shivakumar Raman \\
    Department of Industrial and Systems Engineering \\
    University of Oklahoma \\
    Norman, Oklahoma-73071 \\
    \texttt{raman@ou.edu} \\
}

\begin{document}
\maketitle

\begin{abstract}

Bearings play an integral role in ensuring the reliability and efficiency of rotating machinery – reducing friction and handling critical loads. Bearing failures that constitute up to 90\% of mechanical faults highlight the imperative need for reliable condition monitoring and fault detection. This study proposes a multimodal bearing fault classification approach that relies on vibration and motor phase current signals within a one-dimensional convolutional neural network (1D CNN) framework. The method fuses features from multiple signals to enhance the accuracy of fault detection. Under the baseline condition (1,500 rpm, 0.7 Nm load torque, and 1,000 N radial force), the model reaches an accuracy of 96\% with addition of L2 regularization. This represents a notable improvement of 2 percentage points compared to the non-regularized model. In addition, the model demonstrates robust performance across three distinct operating conditions by employing transfer learning (TL) strategies. Among the tested TL variants, the approach that preserves parameters up to the first max-pool layer and then adjusts subsequent layers achieves the highest performance. While this approach attains excellent accuracy across varied conditions, it requires more computational time due to its greater number of trainable parameters. To address resource constraints, less computationally intensive models offer feasible trade-offs, albeit at a slight accuracy cost. Overall, this multimodal 1D CNN framework with late fusion and TL strategies lays a foundation for more accurate, adaptable, and efficient bearing fault classification in industrial environments with variable operating conditions.

\end{abstract}

\keywords{Bearing Fault Detection\and 1D CNN\and  Transfer Learning\and Multimodal Fusion\and L2 Regularization}


\section{Introduction}\label{sec1}

Rotating machinery, comprising of turbines, motors, pumps, and compressors, is essential for a variety of applications in the manufacturing, transportation and power generating industry~\cite{gangsar2022review}.  The efficiency and reliability of these machinery are important to ensuring production and operational safety. At the heart of these systems are bearings – a key tribological part – used to minimize friction and wear between rotating and stationary parts~\cite{bhuiyan2023deep}. Bearings also sustain axial and radial loads while minimizing resistance, which reduces energy consumption. Rotating machinery relies heavily on bearing health to function properly. Bearings are more likely to fail owing to extended exposure to harsh working conditions (high temperature, high rotational speed, and high load) than other mechanical elements. According to statistics, based on the type and size of the machine, bearing failure accounts for 40\% to 90\% of the mechanical faults~\cite{wang2021vibro}.  Therefore, bearing maintenance and condition monitoring are of paramount importance to ensure that rotating machinery operates continuously and efficiently. To minimize maintenance costs and avert potential calamities, it is pertinent to diagnose faults promptly and accurately before extensive damage occurs.  

Monitoring the condition of bearings and accurately predicting and classifying faults is crucial, particularly in engineering applications that demand high reliability, such as helicopters, aero-engines, high-speed trains, and wind turbines. With the advent of modern sensor technology, it has become possible in recent years to capture signals containing key information about the health status of mechanical systems. In fact, vibration, motor current, acoustic, and thermal imaging data are commonly utilized for classification of bearing faults. Leveraging these signals, data-driven methods have demonstrated effectiveness to accurately diagnose faults in bearings~\cite{kannan2024review}. 

Vibration analysis provides a robust and reliable means of monitoring machine operations. Its non-invasive nature and capacity for ongoing, uninterrupted monitoring are making it more widely adopted in the industry~\cite{mehta2021machine}. More than 70\% of mechanical failures are indicated by vibration, according to statistical data. Vibration signals from a machine provide valuable insights into its condition. Healthy machines maintain constant, low-amplitude vibrations, in contrast to faulty machines, which produce variable vibrations over time. With recent progress in artificial intelligence (AI) and the emergence of low-cost vibration sensors, researchers are increasingly focused on developing efficient machine fault diagnosis techniques based on extensive vibration data. Toma et al. (2021) used a convolutional neural networks (CNN) based classifier model for bearing fault classification using vibration data. In this work, the authors transformed the 1D vibration data to 2D images using continuous wavelet transform for effective training of the CNN model~\cite{toma2021bearing}. In addition, Magar et al. (2021), implemented deep CNN for classification of bearing faults using vibration data. The authors incorporated statistical feature extraction from raw signals such as mean, variance and kurtosis to improve classification performance. Their model was validated using Case Western Reserve University (CWRU) and Paderborn University (PU) bearing dataset~\cite{magar2021faultnet}. Similarly, Altaf et al. (2022), developed a novel approach to classify bearing faults by extracting average, kurtosis, skewness and root mean square in both time and frequency domain. The extracted features were tested using K-Nearest Neighbor (KNN) and Support Vector Machine (SVM) algorithm to classify between ball, inner race, and outer race faults~\cite{altaf2022statistical}. On the other hand, Karpat et al. (2021) proposed 1D CNN based architecture using vibration data to diagnose rolling bearing faults. The proposed model’s performance was assessed under variable operating conditions. The authors achieved an accuracy of 93.97\% for classifying healthy state, inner raceway and outer raceway faults, using the proposed 1D CNN model~\cite{karpat2021cnn}. 

On the other hand, the use of current signals for condition monitoring is noninvasive and inexpensive, offering economic efficiency and simplicity in implementation. Moreover, current signals are resistant to environmental noise. Motor current signals offer complementary information that can detect electrical anomalies in bearings easily, using current transducers, without needing to install sensors around the bearing~\cite{hoang2019motor}. In recent years, there have been various studies that have used motor current signals for bearing fault classification using machine learning (ML) and deep learning (DL) techniques. For instance, Dhomad and Jaber (2020), proposed artificial neural network (ANN) using extracted features from time domain of current signals, such as, root mean square, kurtoses, skewness, and standard deviation and crest factor, to classify between healthy, inner race faults, and outer race faults bearings~\cite{dhomad2020bearing}. In contrast, Wagner and Sommer (2020), used 1D-CNN-LSTM and weighted ensemble learning using multiple motor phase current signals to differentiate between healthy, inner race faults, and outer race faults. Experimental results portray high accuracy of 98.93\%, indicating the effectiveness of the proposed model~\cite{wagner2020bearing}. Likewise, Toma et al. (2020) proposed a hybrid model utilizing motor current signals with statistical feature extraction, genetic algorithm for feature selection, and machine learning classifiers, namely, KNN, decision tree and random forest, for bearing fault classification, achieving accuracy over 97\%~\cite{toma2021bearing}.

Fault classification based on single modal measurements often fails to take into account the complex and comprehensive nature of the faults in bearings. In addition, single modal sensor may produce missed detection and is susceptible to external interferences and noise, therefore, such methods may not be adequate for making a reliable decision on bearing fault diagnosis~\cite{wang2022rolling}. Research has demonstrated that fault classification with multi-sensor signals can achieve better results in contrast to single-sensor signals since multi-sensor signals usually contain complementary fault features~\cite{chen2017multisensor}. Narwade et al. (2013) in their research focused on fault detection of bearing damages in induction motors, using both vibration and current signals~\cite{narwade2013fault}. Additionally, the work by Wang et al. (2021), presented a method of bearing fault diagnosis integrating multi-modal sensor signals - vibration and acoustic data, deploying 1D CNN with fusion algorithm. The signals are concurrently collected by an accelerometer (vibration) and microphone (acoustic), and the features are extracted at the stage of 1D CNN based feature extraction. After feature extraction, the extracted features are merged at the fusion stage, thereby using a softmax classifier to differentiate between types of bearing faults. This novel method has demonstrated superior results under different noisy environments over existing algorithms based of single modal signals~\cite{wang2021vibro}. Furthermore, Pacheco-Cherrez et al. (2022), in their paper, compared different machine learning classification methods by extracting time - and frequency-domain features from both vibration and current signal data, achieving accuracy of more than 96\%. This advantage of multi-sensor fusion, i.e., multi-modal data is not limited to fault diagnosis alone but has been successfully applied across a range of fields~\cite{pacheco2022bearing}. In particular, multi-sensor fusion has been widely adopted in areas such as robotics and biomedical systems (Luo et al., 2022), image processing (Nie et al., 2021), and manufacturing process monitoring (Kong, 2020).

Deep Learning (DL), one of the sub-areas of machine learning have further enhanced the integration of multi-sensor signals fusion. DL can extract the features of input samples layer by layer using deep networks, and automatically extract features using nonlinear activation function at each layer~\cite{wang2020automatic}. Consequently, DL eliminates the need for manual extraction and dependance on human intervention. DL advancements, notably convolutional neural networks (CNNs), have transformed signal processing and classification. CNNs are well-known for their ability to automatically train hierarchical feature representations from raw data, making them an effective tool for studying complicated signals in a variety of fields including image identification, computer vision and natural language processing~\cite{taye2023understanding}.  Their capacity to capture spatial dependencies via convolutional layers has proven extremely useful in tasks requiring pattern recognition and defect detection. While CNNs are commonly used for image data, one-dimensional convolutional neural networks (1D CNN), have been specifically designed for time series data, such as vibration and motor current signals. 1D CNN excel in acquiring temporal patterns and correlations within sequential data, making them ideal for fault classification in industrial applications – varying loads and speeds. By applying convolutional filters directly to time-series data, 1D CNNs can extract relevant features for detecting subtle variations in signal behavior, leading to more accurate and reliable fault diagnosis~\cite{kiranyaz2021survey}. The studies by Ince et al. (2016), Kiranyaz et al. (2018), and Eren et al. (2019) implemented 1D CNN models for bearing fault detection and was successful in achieving high detection accuracy~\cite{ince2016realtime, kiranyaz2018realtime, eren2019generic}. 
    
However, applying multimodal 1D CNN models across different operating conditions (which is common for bearing fault classification) can still pose challenges. This is where transfer learning (TL) comes into play. TL is a machine learning technique that involves using a pre-trained model one task (i.e., source domain) to improve the performance of a related task (i.e., target domain). In TL, the pre-trained model's knowledge is transferred to a new model that is being trained on a separate but related task~\cite{ahsan2023monkeypox}. By allowing the model to transfer knowledge learnt from one operating condition to another, transfer learning tackles the issue of variable signal characteristics and enables greater generalization across diverse settings. In addition to boosting model performance, transfer learning considerably reduces training time under new operating conditions by reusing previously learned features, reducing the need to train the model from scratch. Furthermore, this strategy lowers the number of trainable parameters by freezing the CNN's base layers, which are responsible for feature extraction, leaving just the classification layers to be fine-tuned for the current task~\cite{weiss2016transfer}. This leads to faster convergence and cheaper computing costs, making fault classification more efficient in various operating conditions with limited data~\cite{pan2009transfer}. 

None of the previous research has applied 1D CNN with feature-level fusion (late fusion) for vibration, and two motor phase current data, along with transfer learning for variable operating conditions. Thus, it was imperative to develop a model that focuses on enhancing accuracy, robustness and generalization of fault classification of bearings. Taking these opportunities into account, our study focuses on multimodal analysis using raw signals from two motor phase current signals, and vibration data to assist in fault classification of bearings using deep learning approach. A novel multimodal 1D Convolution Neural Network (1D CNN) with late fusion is proposed, i.e., features are extracted separately from two phase current signals, and vibration signals using CNNs, followed by merging of these features into a single feature before the final classification. This approach ensures strength from both the data modalities, hence, taking advantage of complimentary information both possesses for more accurate bearing fault classification. 
    
First, the proposed 1D-CNN model with late fusion (Baseline Model) is validated using Paderborn University (PU) Bearing Dataset with the first operating condition of 1,500 rpm, load torque of 0.7 Nm, and a 1,000 N radial force on the bearing, henceforth referred as Baseline Condition. Subsequently, the Baseline Model was fine-tuned using transfer learning to adapt the model for the remaining three operating conditions. 
Our contribution in this paper are as follows:
\begin{itemize}
    \item Development of a Baseline Model, 1D CNN model using multimodal data (vibration and two motor phase current signals) with late fusion, incorporating two convolutional layers followed by max-pooling and a fully connected layer for fault classification. The model's performance was evaluated with and without L2 regularization (to tackle overfitting issues) to assess its impact on accuracy, precision, recall, and F1-score.

    \item Comprehensive model evaluation across multiple data splits, comparing the performance of Baseline Model and transfer learning models under different training-to-testing ratios (80/20, 70/30, 60/40). The models were tested using a range of epochs (10 and 50), and their performance was measured in terms of accuracy, precision, recall, and F1-score.
    \item Implementation of transfer learning models for fault classification across different operating conditions (3 operating conditions). Three transfer learning models were trained for each of the operating conditions, utilizing pre-trained weights from the Baseline Condition and fine-tuning the models for new environments. The architecture was tested for transfer learning models to verify its generalization capabilities across different operating conditions. Accuracy, precision, recall, F1-score, trainable parameters, and total training time required were computed for each of the transfer learning model.  
\end{itemize}
\section{Methodology}
\subsection{Dataset Description}
\begin{figure}[h!]
    \centering
    \includegraphics[width=0.8\linewidth]{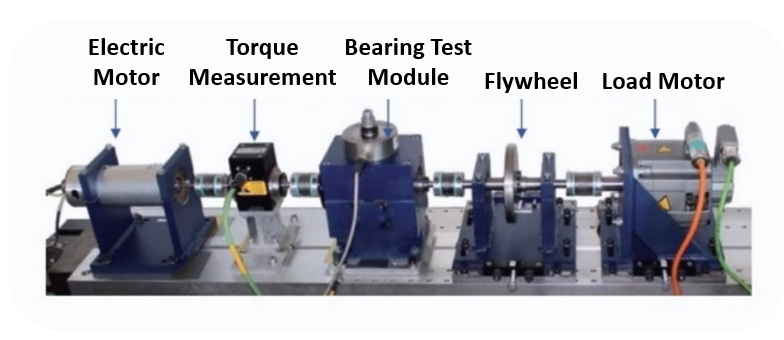}
    \caption{Testbed for Paderborn dataset.}
    \label{fig:paderborn}
\end{figure}

The Paderborn University (PU) Dataset is a benchmark data set for condition monitoring of rolling bearings and is collected from the work by Lessmeier et al. (2016)~\cite{lessmeier2016benchmark}. It serves as a diagnostic technique for the fault classification of bearings. The dataset consists of two motor phase current signals and vibration signals are generated from 32 bearings (Figure~\ref{fig:paderborn}). Out of 32 bearings, 6 are healthy, 12 bearings have artificial damage (created by drilling, EDM and electric engraving machine), 14 bearings are naturally damaged (produced by accelerated lifetime tests). The artificial defects are formed on both inner and outer race. The bearing samples are distributed into 3 classes: healthy, inner race fault, and outer race fault. Based on this classification, there are 6 healthy bearings, 11 inner race fault bearings, and 12 outer race fault bearings. Since the other 3 bearings have multiple faults in them, they were omitted from the dataset for our study, leaving only the bearings that can be classified clearly. 
The dataset consists of four operating conditions. The rotational speed of the drive system, the radial force onto the test bearing and the load torque in the drive train are the primary operation parameters. All three parameters are kept constant for each experiment. For our study, one of the operating condition is used as the baseline condition, where the test rig runs at n = 1,500 rpm with a load torque of M = 0.7 Nm, and a radial force on the bearing of F = 1,000 N. All other operating conditions, including the baseline condition, are given in Table~\ref{tab:experimental_settings}. For each operating condition, each bearing is measured 20 times to generate 20 signals for 4 seconds with a sampling frequency of 64 kHz (for both motor phase current signals and vibration signals). That means there are 256,000 data points in a signal. 
\begin{table}[h!]
\centering
\caption{Four Operating Conditions in the Paderborn University Dataset.}
\begin{tabular}{@{}ccccc@{}}
\toprule
No. & Rotational Speed [rpm] & Load Torque [Nm] & Radial Force [N] & Name of Setting \\ 
\midrule
0 (Baseline Condition) & 1500 & 0.7 & 1000 & N15\_M07\_F10 \\ 
1 & 900 & 0.7 & 1000 & N09\_M07\_F10 \\ 
2 & 1500 & 0.1 & 1000 & N15\_M01\_F10 \\ 
3 & 1500 & 0.7 & 400 & N15\_M07\_F04 \\ 
\bottomrule
\end{tabular}
\label{tab:experimental_settings}
\end{table}

\subsection{Data Segmentation with Overlap Sliding Window}
\begin{figure}[h!]
    \centering
    \includegraphics[width=\linewidth]{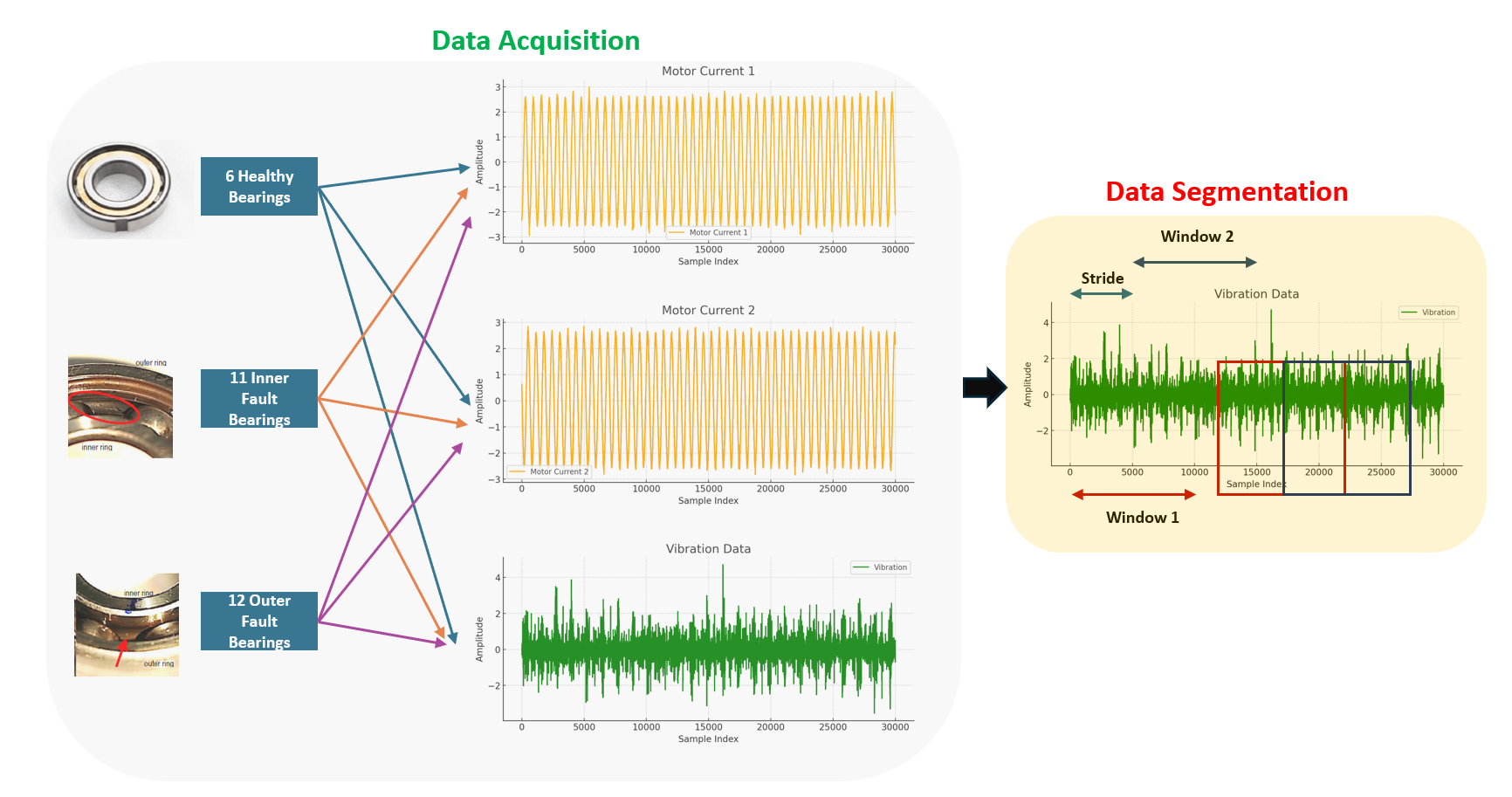}
    \caption{Overlap Sliding Window Creation.}
    \label{fig:slide}
\end{figure}

All the three columns of data (Motor Phase Current 1, Motor Phase Current 2, and Vibration data) are standardized first, to transform the data so that it has a mean of 0 and a standard deviation of 1. After standardization of data, to obtain more samples from this long time series data for the experiments, a data segmentation method with the overlap sampling window operation was implemented (Figure~\ref{fig:slide}). A window length of 10000 rows were used, with a stride length of 5000. Each short segment with same length of data (samples) falls under a bearing condition, i.e., healthy, inner fault or outer fault bearing, based on the original time series data it belonged to. 
\subsection{Proposed 1D CNN Structure with Late Fusion}
1D CNN is widely used for processing time-domain signals, i.e., time series data~\cite{wang2022rolling}. The 1D CNN is a classic deep feedback neural network, where the internal features of the input data are extracted based on the convolution operation. The convolution operation on the input data is only performed on one dimension in 1D CNN, unlike 2D CNN models~\cite{wang2022rolling}. In the CNN structure, feature extraction is a critical component~\cite{ahsan2024enhancing}. More importantly, 1D CNN has powerful feature extraction capability – automatically extracting the non-linear patterns within the raw data by employing alternating convolution layer and pooling layer, finally, the adaptive learning in accomplished in the fully connected layer. Essentially, 1D CNN allows to bypass the manual feature extraction process of the traditional algorithms, facilitating end to end processing~\cite{wang2021vibro}.

\begin{figure}[h!]
    \centering
    \includegraphics[width=\linewidth]{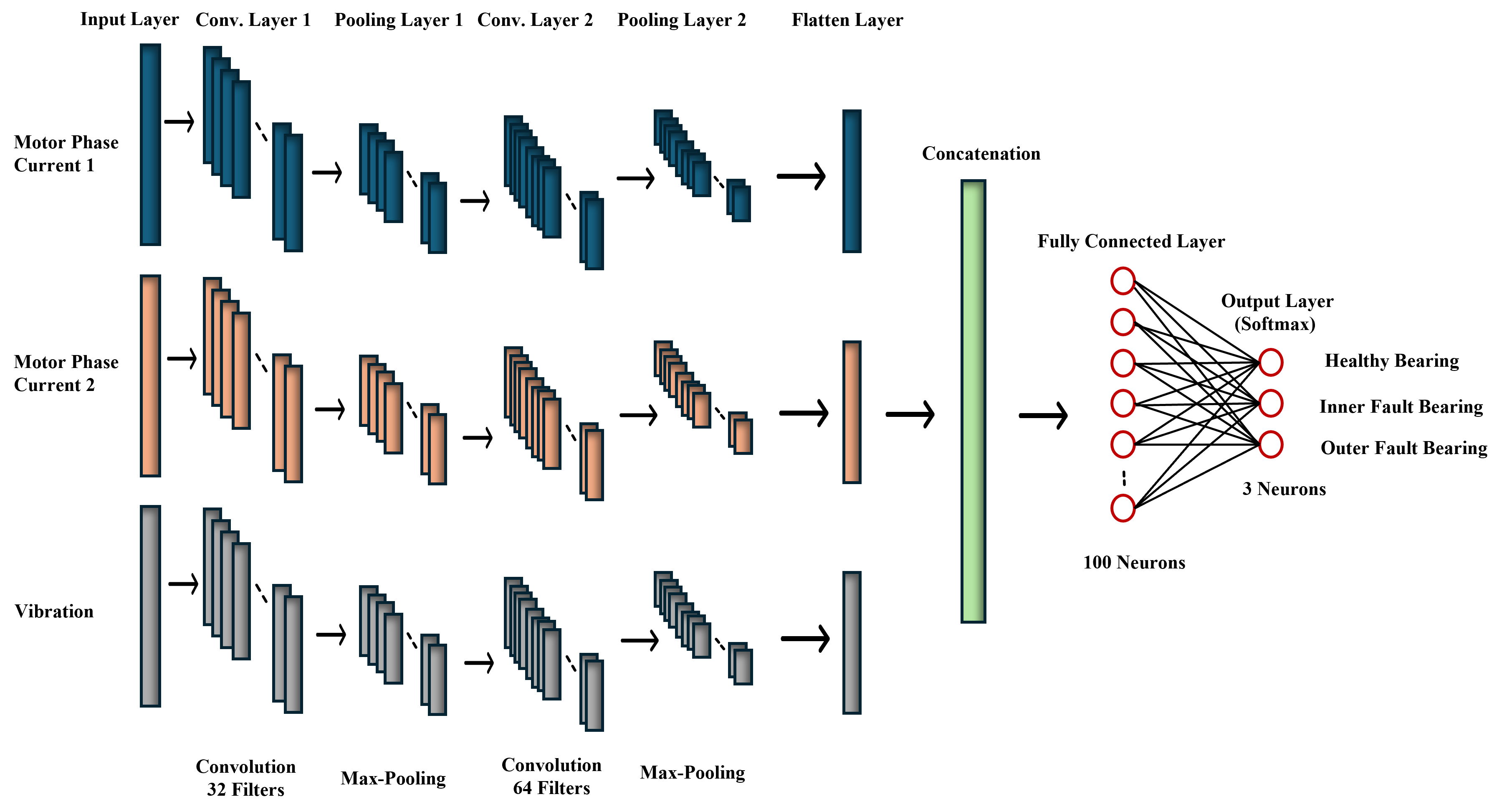}
    \caption{The Proposed Multimodal 1D CNN with Late Fusion.}
    \label{fig:late}
\end{figure}

Figure~\ref{fig:late}., depicts the structure of the proposed 1D CNN model with late fusion for the three input signals, partially adapted from the work by Wang et al. (2021)~\cite{wang2021vibro}. The model architecture consists of two Convolution Layers and two Max-pooling Layers for each of the input signals. The first Convolution Layer utilizes 32 filters with a size of 3, whereas the second Convolution Layer applies 64 filters with a size of 3. 

Each of the Convolution Layer are followed by a Max-pooling Layers, to down sample the feature maps generated from the Convolution Layer, in turn reducing the dimensionality of the data while retaining the key features. These sequence of operations are repeated for all three input branches. Subsequently, the output of the max-pooling layer in each branch is flattened in the Flatten Layer, followed by a Concatenation Layer (fusion layer), where features from all three branches are fused into a single, unified feature representation. Next, these combined features are fed into a fully connected layer with 100 neurons, activated by ReLu function. Finally, the features are imported to the Softmax Layer (Output Layer) with 3 neurons, which is responsible for classification of bearings (healthy bearings, inner fault bearings, and outer fault bearings). The parameters involved in the proposed methodology is portrayed in Table~\ref{tab:cnn_methodology}. 

The model is compiled with the Adam optimizer and sparse categorical cross-entropy loss for optimal classification of bearings. 
\begin{table}[h!]
\centering
\caption{Details of the proposed 1D CNN methodology}
\begin{tabular}{@{}ccccc@{}}
\toprule
No. & Layer Type                          & Output Shape & Number of Filters & Number of Neurons \\ 
\midrule
1  & Input Layer (Phase Current 1)       & 10000, 1     & -                 & -                 \\
2  & Input Layer (Phase Current 2)       & 10000, 1     & -                 & -                 \\
3  & Input Layer (Vibration)             & 10000, 1     & -                 & -                 \\
4  & Convolution Layer 1 (Phase Current 1) & 9998, 32     & 32                & -                 \\
5  & Max-Pooling 1 (Phase Current 1)     & 4999, 32     & -                 & -                 \\
6  & Convolution Layer 1 (Phase Current 2) & 9998, 32     & 32                & -                 \\
7  & Max-Pooling 1 (Phase Current 2)     & 4999, 32     & -                 & -                 \\
8  & Convolution Layer 1 (Vibration)     & 9998, 32     & 32                & -                 \\
9  & Max-Pooling 1 (Vibration)           & 4999, 32     & -                 & -                 \\
10 & Convolution Layer 2 (Phase Current 1) & 4997, 64     & 64                & -                 \\
11 & Max-Pooling 2 (Phase Current 1)     & 2498, 64     & -                 & -                 \\
12 & Convolution Layer 2 (Phase Current 2) & 4997, 64     & 64                & -                 \\
13 & Max-Pooling 2 (Phase Current 2)     & 2498, 64     & -                 & -                 \\
14 & Convolution Layer 2 (Vibration)     & 4997, 64     & 64                & -                 \\
15 & Max-Pooling 2 (Vibration)           & 2498, 64     & -                 & -                 \\
16 & Flatten Layer (Phase Current 1)     & 159872       & -                 & -                 \\
17 & Flatten Layer (Phase Current 2)     & 159872       & -                 & -                 \\
18 & Flatten Layer (Vibration)           & 159872       & -                 & -                 \\
19 & Concatenate                         & 479616       & -                 & -                 \\
20 & Fully Connected Layer               & 100          & -                 & 100               \\
21 & Softmax Layer (Output)              & 3            & -                 & 3                 \\
\bottomrule
\end{tabular}
\label{tab:cnn_methodology}
\end{table}

\subsection{Transfer Learning Strategies}
In conventional deep learning methods, the training and testing set are necessary to be independent and identically distributed (Yang et al., 2023). However, with variable working conditions, the training and the test set have a domain shift, thereby, restricting the performance of deep learning algorithms. Transfer Learning (TL) is known for utilizing shared features between two different but related domains. In particular, parameter and feature transfer method from TL, have been extensively used in the field of fault diagnosis (Yang et al., 2023). 

In parameter transfer, the emphasis is on modification of model’s weight. Firstly, the model is pre-trained with the source domain, followed by fine-tuning of the model parameters from the target domain, to enable bearing fault classification under variable working conditions. 

In this study, the transfer learning strategies were partially adapted from the work by Liu et al. (2021)~\cite{li2021central}. For each of the TL models (1, 2 and 3), the baseline operating condition mentioned in Table~\ref{tab:experimental_settings}, represents as the source domain from which the TL models are initially trained. The three TL models are developed as follows:

\begin{itemize}
    \item Model 1 was developed by applying the pre-trained model for weight initialization. The weights of the pre-trained model were frozen for all the layers except the fully connected layer for all input signals.  The fine-tuning was performed on the fully connected layer to adapt the model to the target domain data (Figure~\ref{fig:model1}).

    \begin{figure}[h!]
    \centering
    \includegraphics[width=\linewidth]{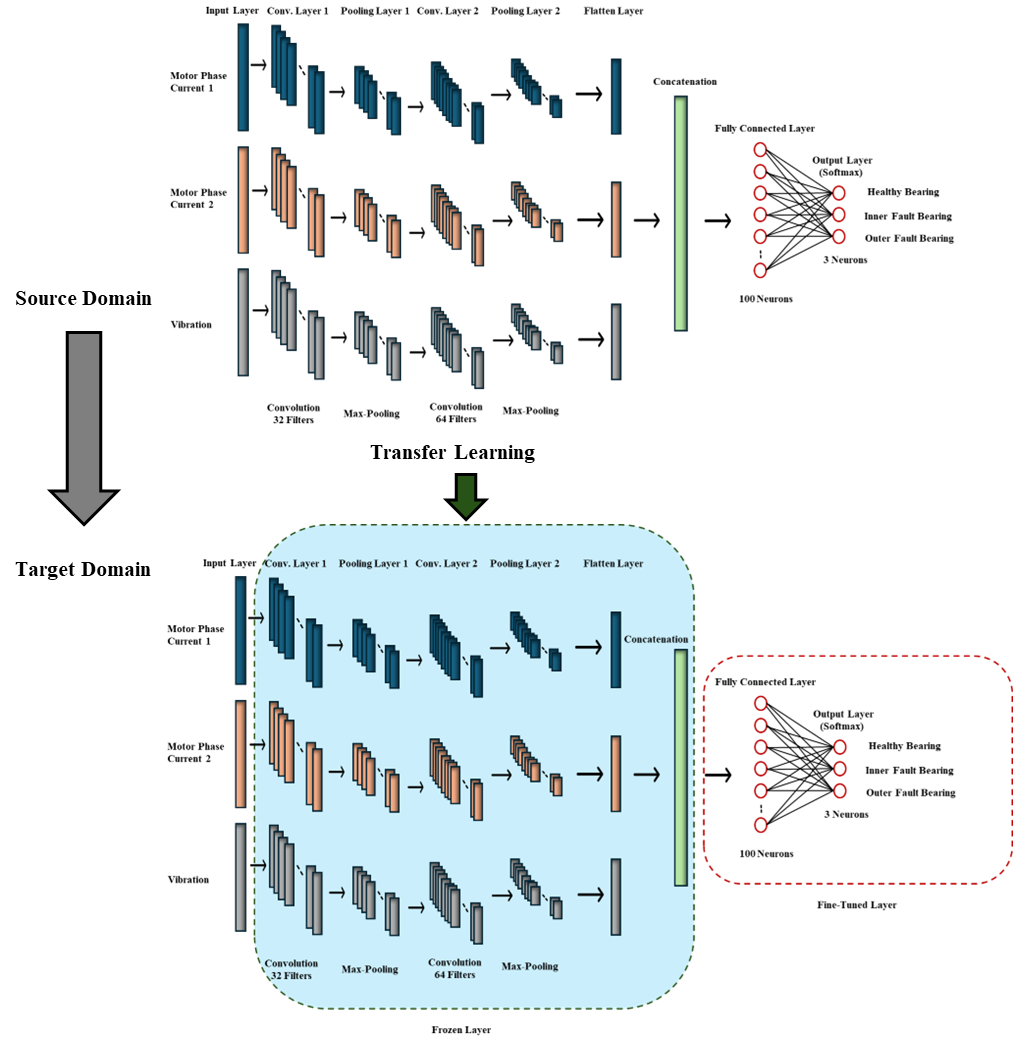}
    \caption{Framework of proposed Transfer Learning (Model 1).}
    \label{fig:model1}
\end{figure}

    \item Model 2 was developed by freezing the first Convolution Layer and Max-pooling Layer for each of the input signals. The fine-tuning was conducted on all the layers after the first Convolution Layer and Max-pooling Layer to learn the parameters of the target domain data (See Appendix~\ref{sec:appendix}, Figure~\ref{fig:model2}).
    \item Model 3 was constructed by freezing all the layers before the fully connected layers for all the input signals, while replacing the original fully connected layer with the new ones to adapt to the target domain. In other words, the newly added fully connected layer adapts specifically to the target domain data, whereas the knowledge from the source domain are maintained due to frozen layers (See Appendix~\ref{sec:appendix}, Figure~\ref{fig:model2}). . 

\end{itemize}

\subsection{Experimental Setup}
\subsubsection{Baseline Operating Condition}
The entire experiment was initially implemented using an office-grade laptop running a Windows 10 operating system, equipped with an Intel Core i7-10750 H processor and 16 GB of RAM. To leverage high-performance computing, further experiments were conducted on Google Colab, utilizing an NVIDIA A100 GPU. Table~\ref{tab:experimental_dataset} summarizes the dataset used in this study. As indicated in the table, a total of 29783 samples of 10000 length were generated for the experiments using the sliding window technique. These generated samples are from 6 healthy bearings, 11 inner fault bearings and 12 outer fault bearings. Three different training set / testing set ratios were used, namely, 80/20, 70/30, and 60/40. 

\begin{table}[h!]
\centering
\caption{Details of the experimental dataset (PU)}
\label{tab:experimental_dataset}
\begin{tabular}{@{}lcccc@{}}
\toprule
\textbf{Train/Test Split} & \textbf{Label}     & \textbf{Train Set} & \textbf{Test Set} & \textbf{Total} \\ \midrule
\multirow{4}{*}{80/20}    & Healthy            & 4929              & 1231             & 6160           \\
                          & Inner Fault        & 9037              & 2259             & 11296          \\
                          & Outer Fault        & 9858              & 2464             & 12322          \\
                          & \textbf{Total}     & \textbf{23825}    & \textbf{5957}    & \textbf{29783} \\ \midrule
\multirow{4}{*}{70/30}    & Healthy            & 4312              & 1848             & 6160           \\
                          & Inner Fault        & 7907              & 3389             & 11296          \\
                          & Outer Fault        & 8625              & 3697             & 12322          \\
                          & \textbf{Total}     & \textbf{20848}    & \textbf{8935}    & \textbf{29783} \\ \midrule
\multirow{4}{*}{60/40}    & Healthy            & 3696              & 2464             & 6160           \\
                          & Inner Fault        & 6777              & 4519             & 11296          \\
                          & Outer Fault        & 7393              & 4929             & 12322          \\
                          & \textbf{Total}     & \textbf{17869}    & \textbf{11914}   & \textbf{29783} \\ \bottomrule
\end{tabular}
\end{table}

For the respective training sets, 5-folds cross validation was implemented to monitor any overfitting before going into the testing set. For each of the training set / testing set split, results were generated for epochs of 10 and 50 with a batch size of 32, both with and without L2 regularization. For the L2 regularization, a regularization strength (lambda) of 0.01was utilized, where lambda controls the penalty applied to the model’s weight. Furthermore, for the classification, Adam optimizers with a constant learning rate of 0.001, and sparse cross entropy loss were applied. L2 regularization is a crucial technique in CNNs, that is instrumental in reducing overfitting by penalizing large weights in the model. Moreover, by penalizing large weights, it prevents the model from fitting noises in the training set – a common problem in deep learning networks with many parameters. In a nutshell, L2 regularization helps achieve higher accuracy and reduce overfitting in the model. 

\subsubsection{For Other Operating Conditions}
For all three transfer learning models, we allocated 60\% of the data for training, 20\% of the data for validation, and 20\% of the data for testing. The models were trained with 50 epochs and a batch size of 32. Like the baseline condition, sparse cross entropy loss was implemented, along with Adam optimizer, however, a lower learning rate of 1e-5 is applied. Fine-tuning with lower learning rate allows for preserving useful features from the source domain by preventing large weight updates, while gradually adapting to target domain. 

\subsubsection{Performance Evaluation}
The performance assessment for proposed CNN model for the baseline conditions, and the transfer learning models was conducted using various metrics, as described below~\cite{ahsan2023monkeypox}: 
\begin{equation}\label{eq1}
    Accuracy = \frac{\sum_{i=1}^{C} TP_i}{\sum_{i=1}^{C} (TP_i + FP_i + FN_i)}
\end{equation}

\begin{equation}\label{eq2}
    Precision_i = \frac{TP_i}{TP_i + FP_i}
\end{equation}

\begin{equation}\label{eq3}
    Recall_i = \frac{TP_i}{TP_i + FN_i}
\end{equation}

\begin{equation}\label{eq4}
    F1-Score_i = 2 \times \frac{\textrm{Precision}_i \times \textrm{Recall}_i}{\textrm{Precision}_i + \textrm{Recall}_i}
\end{equation}

\begin{equation}\label{eq5}
    CI = \hat{p} \pm Z_{\alpha/2} \times \sqrt{\frac{\hat{p} (1 - \hat{p})}{n}}
\end{equation}

Here,

$C$ = Number of classes (e.g., Healthy, Inner fault, Outer fault)

$TP_i$ = Number of true positives for class $i$

$FP_i$ = Number of false positives for class $i$

$FN_i$ = Number of false negatives for class $i$

$\hat{p}$ = Sample proportion (accuracy in this case)

$Z_{\alpha/2}$ = Z-value from the standard normal distribution (1.96 for a 95\% confidence level)

$n$ = Sample size

ROC curves illustrate a model’s performance, plotting the true positive rate against the false positive rate at different threshold settings. It is useful to assess the trade-offs between sensitivity and specificity. The area under the ROC curve (AUC) is a popularly used single-value metric that encapsulates the overall performance of a classifier. 

For the transfer learning models, two other parameters are tracked, i.e., the trainable parameters, and the total processing time required for training. These metrics are computed to validate the importance of transfer learning for variable operating conditions, compared to using the baseline operating condition to train the model, and testing on the unseen test set, which would be three operating conditions, respectively.

\section{Results}
\subsection{Baseline Model}

Table~\ref{tab:baseline_performance} presents the results of the proposed 1D CNN model with late fusion for the baseline operating condition, for varied test set size (20\%, 30\%, and 40\%), and number of epochs (10 and 50). The overall accuracy, precision, recall and F1-score were computed for each of the test set size and epochs respectively. It is observed that test set size 30\% with 50 epochs yield the best results with accuracy of 0.94, precision of 0.94, recall of 0.94, and F1-score of 0.94. Moreover, the results from 20\% test set size were close to the best performance, with a score of 0.93 across all metrics, for both 10 and 50 epochs. On the other hand, test data size of 40\% displayed the lowest performance for 50 epochs, with a score of 0.87 across all metrics. However, with L2 regularization (as shown in Table~\ref{tab:baseline_l2}) , the best performance was observed for the test data size of 20\% and 50 epochs, with a value of 0.96 across all the metrics. This is an improvement from 0.93 for the similar setting, just without L2 regularization. Interestingly, the best performing setting without L2 regularization, has a deterioration is performance with L2 regularization with accuracy, recall and F1 score reducing from 0.94 to 0.89, respectively; moreover, precision decreasing from 0.94 to 0.90. 
\begin{table}[h!]
\centering
\caption{The performance of the baseline Operating Condition N15\_M07\_F10}
\label{tab:baseline_performance}
\begin{tabular}{@{}lcccccc@{}}
\toprule
\textbf{Test Data} & \textbf{Epochs} & \textbf{Accuracy} & \textbf{Precision} & \textbf{Recall} & \textbf{F1-Score} \\ \midrule
20\%               & 10              & 0.93              & 0.93               & 0.93            & 0.93              \\
                   & 50              & 0.93              & 0.93               & 0.93            & 0.93              \\ \midrule
30\%               & 10              & 0.93              & 0.93               & 0.93            & 0.93              \\
                   & 50              & 0.94              & 0.94               & 0.94            & 0.94              \\ \midrule
40\%               & 10              & 0.90              & 0.90               & 0.90            & 0.90              \\
                   & 50              & 0.87              & 0.87               & 0.87            & 0.87              \\ \bottomrule
\end{tabular}
\end{table}
\begin{table}[h!]
\centering
\caption{The performance of the baseline Operating Condition N15\_M07\_F10 with L2 Regularization}
\label{tab:baseline_l2}
\begin{tabular}{@{}lcccccc@{}}
\toprule
\textbf{Test Data} & \textbf{Epochs} & \textbf{Accuracy} & \textbf{Precision} & \textbf{Recall} & \textbf{F1-Score} \\ \midrule
20\%               & 10              & 0.92              & 0.93               & 0.92            & 0.92              \\
                   & 50              & 0.96              & 0.96               & 0.96            & 0.96              \\ \midrule
30\%               & 10              & 0.93              & 0.93               & 0.93            & 0.93              \\
                   & 50              & 0.89              & 0.90               & 0.89            & 0.89              \\ \midrule
40\%               & 10              & 0.93              & 0.94               & 0.93            & 0.94              \\
                   & 50              & 0.96              & 0.96               & 0.96            & 0.96              \\ \bottomrule
\end{tabular}
\end{table}

\begin{figure}[h!]
    \centering
    \includegraphics[width=\linewidth]{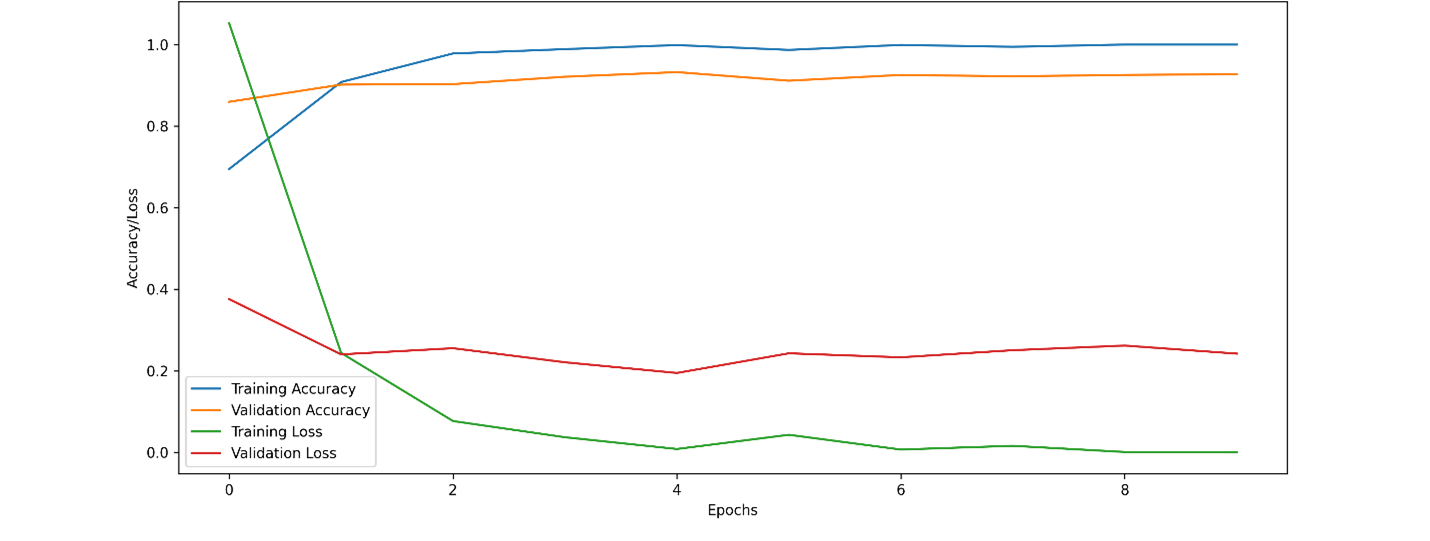}
    \caption{Training accuracy and loss with number of epochs.}
    \label{fig:traintest}
\end{figure}

Figure~\ref{fig:traintest} portrays the model’s performance using Adam optimizer with 10 epochs for the 5th fold of the cross-validation process. Both training and validation loss and accuracy were monitored to understand any under- or over-fitting issues in the training process, and how the model is performing on the unseen validation set before going to the test set. In the same way, Figure 10 presents the graph with L2 regularization, monitoring the model’s training and validation performance across epochs. 

\begin{figure}
    \centering
    \includegraphics[width=.7\linewidth]{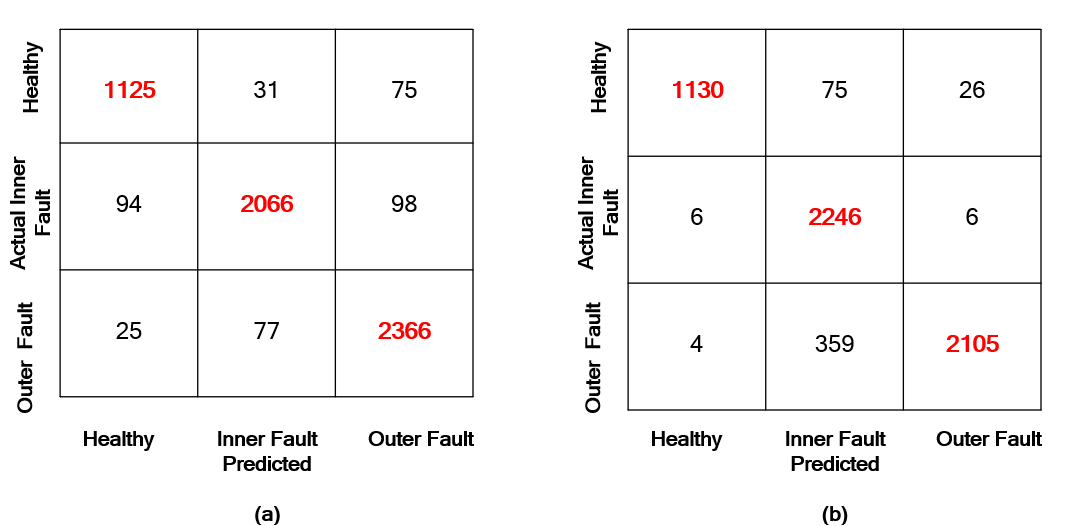}
    \caption{The Confusion matrix of (a) 80/20 Split and (b) with L2 Regularization.}
    \label{fig:con1}
\end{figure}

The confusion matrix for the Baseline Model was depicted in Figure~\ref{fig:con1} (a). Out of the total samples of 5957 in the test set of size 20\% of the total samples of 29783 as shown in Table 3, 5557 samples were correctly predicted. The diagonal in Figure~\ref{fig:con1}(a) shows the correctly classified bearings in the test set, with 1125 samples of Healthy bearings, 2066 Inner Fault bearings, and 2366 Outer Fault bearings classified correctly, respectively. Furthermore, the confusion matrix indicates the misclassified predictions of the proposed model, in other words, called false positives and negatives. For the Healthy bearings, there are 119 (94 and 25) false positives, meaning 94 Inner Fault bearings and 25 Outer Fault bearings were incorrectly classified as healthy. In addition, for Inner and Outer Fault bearings, 108 and 173 false positives, respectively, were reported. Similarly, for Healthy bearings there are 106 false negatives, i.e., Healthy bearings were misclassified as Inner Fault bearings (31), and Outer Fault bearings (75). Likewise, for Inner and Outer Fault bearings, 192 and 102 false negatives were generated in the confusion matrix. 

On the other hand, the confusion matrix with L2 regularization is presented in Figure~\ref{fig:con1} (b). There is a higher number of correctly classified Healthy (1130) and Inner Fault bearing (2246) compared to 1125 and 2066 without L2 regularization. Hover, for Outer Fault bearings the number has lowered from 2366 to 2105. In contrast, false positive for Inner Fault bearings have increased from 108 to 434, and false negative for Outer Fault bearing from 102 to 363. 

\begin{figure}
    \centering
    \includegraphics[width=\linewidth]{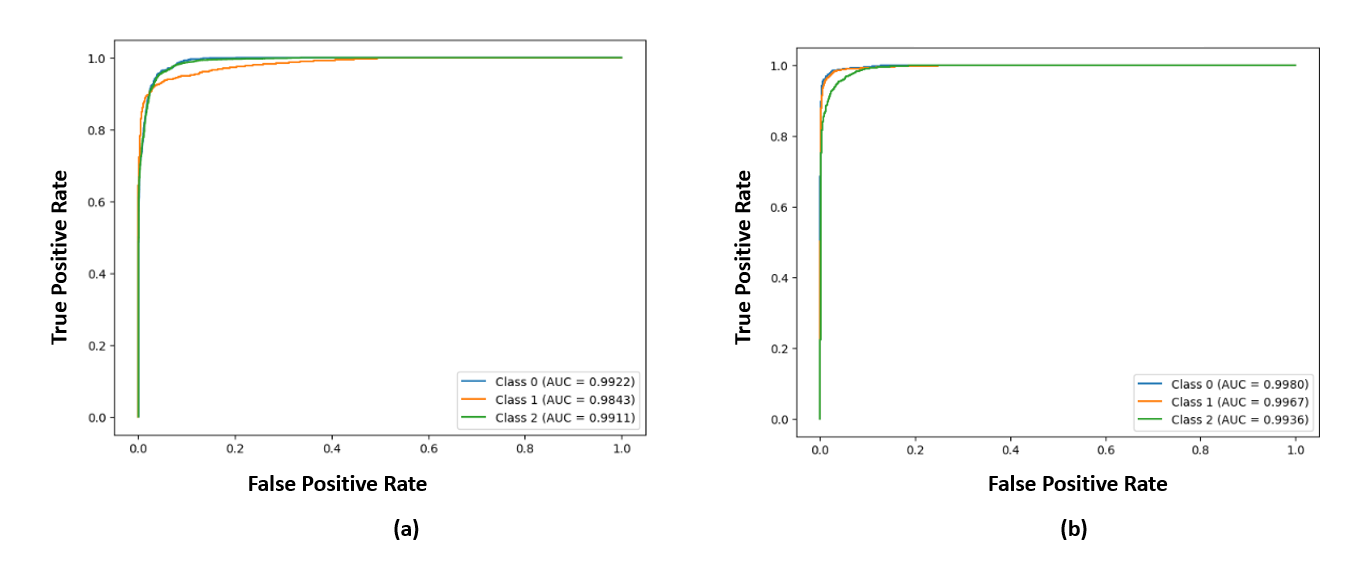}
    \caption{The ROC Curve of (a) 80/20 Split with AUC values and (b) with L2 Regularization for each bearing type.}
    \label{fig:roca}
\end{figure}

Finally, the Receiver Operating Characteristic (ROC) curve with Area Under the Curve (AUC) values is illustrated in Figure~\ref{fig:roca}, for the test set with size 20\%, with True Positive Rate and False Positive Rate as the metrics of evaluation. As demonstrated in Figure~\ref{fig:roca} (a), Class 0 (Healthy bearing) with an AUC value of 0.9922, Class 1 (Inner Fault bearing) with an AUC value of 0.9842, and Class 2 (Outer Fault bearing) with an AUC value of 0.9911, establishes that the model performs exceptionally well (values close to 1) in distinguishing between the three bearing classes.

On the contrary, in Figure~\ref{fig:roca} (b) the AUC values for the bearing types with the L2 regularization have improved for all the classes. Healthy bearings with an AUC value of 0.9980, Inner Fault bearings with an AUC value of 0.9967, and Outer Fault bearings with an AUC value of 0.9936, have exhibited superior results compared to the model without L2 regularization. 

\subsection{Transfer Learning}
Table~\ref{tab:transfer} presents the overall performance of the three operating conditions, which are used individually as the test set, while the baseline operating condition is used as the training set for each scenario. The table depicts that the variable operating conditions deteriorated the performance of the proposed CNN model on each of the metrics. The model performed best on the second operating condition with accuracy of 0.90 ± 0.01, where rotational speed and radial force is consistent with the baseline condition, with the only alteration in load torque value from 0.7 Nm to 0.1 Nm. Conversely, condition 1 performs the worst with accuracy of 0.40 ± 0.01, where parameters load torque and radial force are constant with the baseline condition, with the only change in rotational speed from 1500 rpm to 900 rpm.  

These performances from the variable operating conditions as the test set, with the pretrained model of the baseline operating condition as the train set resulted in poor performance overall, barring from the second operating condition.  
\begin{table}[ht]
    \centering
    \caption{Computational Results of using Three Operating Conditions as Unseen Test Set with Baseline Condition as the Train Set}
    \label{tab:transfer}
    \begin{tabular}{@{}lcccc@{}}
        \toprule
        \textbf{Operating Condition} & \textbf{Accuracy} & \textbf{Precision} & \textbf{Recall} & \textbf{F1 Score} \\
        \midrule
        N09\_M07\_F10 (1) & $0.40 \pm 0.01$ & $0.41 \pm 0.01$ & $0.49 \pm 0.01$ & $0.39 \pm 0.01$ \\
        N15\_M01\_F10 (2) & $0.90 \pm 0.01$ & $0.89 \pm 0.01$ & $0.90 \pm 0.01$ & $0.89 \pm 0.01$ \\
        N15\_M07\_F04 (3) & $0.76 \pm 0.01$ & $0.78 \pm 0.01$ & $0.76 \pm 0.01$ & $0.75 \pm 0.01$ \\
        \bottomrule
    \end{tabular}
    \end{table}

In Table~\ref{tab:tl_models_performance}, we have displayed the performance of the three TL based models on operating condition 1.  The table indicates that there is an improvement in performance for all the metrics in every TL models compared to the results presented in Table 6. The Model 2 showcased the best performance with accuracy of 0.92 ± 0.01, precision of 0.92 ± 0.01, recall of 0.92 ± 0.01, and F1 score of 0.92 ± 0.01 on the test set. Model 1 and 3 delivered similar results of approximate 0.81 ± 0.01 across all metrics.  

\begin{table}[h!]
\centering
\caption{The performance of three TL Models with a Confidence Interval $\alpha = 0.05$ on N09\_M07\_F10 (Operating Condition 1).}
\label{tab:tl_models_performance}
\small
\begin{tabular}{@{}lcccccccc@{}}
\toprule
\textbf{Model} & \multicolumn{4}{c}{\textbf{Train Set}} & \multicolumn{4}{c}{\textbf{Test Set}} \\ 
\cmidrule(lr){2-5} \cmidrule(lr){6-9}
 & \textbf{Accuracy} & \textbf{Precision} & \textbf{Recall} & \textbf{F1 Score} & \textbf{Accuracy} & \textbf{Precision} & \textbf{Recall} & \textbf{F1 Score} \\ 
\midrule
Model 1 & 0.95 ± 0.00 & 0.95 ± 0.00 & 0.95 ± 0.00 & 0.95 ± 0.00 & 0.81 ± 0.01 & 0.81 ± 0.01 & 0.81 ± 0.01 & 0.81 ± 0.01 \\ 
Model 2 & 1.00 ± 0.00 & 1.00 ± 0.00 & 1.00 ± 0.00 & 1.00 ± 0.00 & 0.92 ± 0.01 & 0.92 ± 0.01 & 0.92 ± 0.01 & 0.92 ± 0.01 \\ 
Model 3 & 0.95 ± 0.00 & 0.95 ± 0.00 & 0.95 ± 0.00 & 0.95 ± 0.00 & 0.81 ± 0.01 & 0.82 ± 0.01 & 0.81 ± 0.01 & 0.81 ± 0.01 \\ 
\bottomrule
\end{tabular}
\end{table}
Similarly, Table~\ref{tab:tl_models_performance_op2} illustrated the performance of the TL based models on the 2nd operating condition. Both Model 1 and 2 showed superior performance than the performance presented in Table 6. Model 2 exhibited the best performance with a value of 0.96 ± 0.01 across all the metrics on the test data. In contrast, Model 3 performed worse than the results displayed in Table 6, with an accuracy of 0.81 ± 0.01, precision of 0.82 ± 0.01, recall of 0.81 ± 0.01, and F1 Score of 0.81 ± 0.01. 
\begin{table}[h!]
\centering
\caption{The performance of three TL Models with a confidence interval $\alpha = 0.05$ on N15\_M01\_F10 (Operating Condition 2).}
\label{tab:tl_models_performance_op2}
\small
\begin{tabular}{@{}lcccccccc@{}}
\toprule
\textbf{Model} & \multicolumn{4}{c}{\textbf{Train Set}} & \multicolumn{4}{c}{\textbf{Test Set}} \\ 
\cmidrule(lr){2-5} \cmidrule(lr){6-9}
 & \textbf{Accuracy} & \textbf{Precision} & \textbf{Recall} & \textbf{F1 Score} & \textbf{Accuracy} & \textbf{Precision} & \textbf{Recall} & \textbf{F1 Score} \\ 
\midrule
Model 1 & 1.00 ± 0.00 & 1.00 ± 0.00 & 1.00 ± 0.00 & 1.00 ± 0.00 & 0.92 ± 0.01 & 0.92 ± 0.01 & 0.92 ± 0.01 & 0.92 ± 0.01 \\ 
Model 2 & 1.00 ± 0.00 & 1.00 ± 0.00 & 1.00 ± 0.00 & 1.00 ± 0.00 & 0.96 ± 0.01 & 0.96 ± 0.01 & 0.96 ± 0.01 & 0.96 ± 0.01 \\ 
Model 3 & 1.00 ± 0.00 & 1.00 ± 0.00 & 1.00 ± 0.00 & 1.00 ± 0.00 & 0.81 ± 0.01 & 0.82 ± 0.01 & 0.81 ± 0.01 & 0.81 ± 0.01 \\ 
\bottomrule
\end{tabular}
\end{table}

Likewise, Table~\ref{tab:tl_models_performance_op3} detailed the performance of the TL based model on operating condition 3. Model 2 emerged as the top performer with a value of 0.89 ± 0.01 across all metrics on the test data, whereas Models 1 and 3 displayed same performance with an accuracy of 0.83 ± 0.01, precision of 0.84 ± 0.01, recall of 0.83 ± 0.01, and F1 Score of 0.83 ± 0.01. 

\begin{table}[h!]
\centering
\caption{The performance of three TL Models with a confidence interval $\alpha = 0.05$ on N15\_M07\_F04 (Operating Condition 3).}
\label{tab:tl_models_performance_op3}
\small
\begin{tabular}{@{}lcccccccc@{}}
\toprule
\textbf{Model} & \multicolumn{4}{c}{\textbf{Train Set}} & \multicolumn{4}{c}{\textbf{Test Set}} \\ 
\cmidrule(lr){2-5} \cmidrule(lr){6-9}
 & \textbf{Accuracy} & \textbf{Precision} & \textbf{Recall} & \textbf{F1 Score} & \textbf{Accuracy} & \textbf{Precision} & \textbf{Recall} & \textbf{F1 Score} \\ 
\midrule
Model 1 & 0.94 ± 0.00 & 0.94 ± 0.00 & 0.94 ± 0.00 & 0.94 ± 0.00 & 0.83 ± 0.01 & 0.84 ± 0.01 & 0.83 ± 0.01 & 0.83 ± 0.01 \\ 
Model 2 & 0.99 ± 0.00 & 0.99 ± 0.00 & 0.99 ± 0.00 & 0.99 ± 0.00 & 0.89 ± 0.01 & 0.89 ± 0.01 & 0.89 ± 0.01 & 0.89 ± 0.01 \\ 
Model 3 & 0.94 ± 0.00 & 0.94 ± 0.00 & 0.94 ± 0.00 & 0.94 ± 0.00 & 0.83 ± 0.01 & 0.84 ± 0.01 & 0.83 ± 0.01 & 0.83 ± 0.01 \\ 
\bottomrule
\end{tabular}
\end{table}

In Table~\ref{tab:trainable_params_computation_time}, we have presented the summary of the total trainable parameters, and total process time to train the models that were employed on the baseline model as well as the transfer learning models. For comparison, all the data used in the models were given an equal split, with 60\% assigned for train set, 20\% for validation set, and 20\% for test set across the conditions.  The Model 3 in all the conditions required the least time to train, with a time of 183.03 seconds for condition 1, 182.94 for condition 2, and 184.13 for condition 3. Model 2 took the highest time across operating conditions, since it needed to train more parameters, as illustrated in Figure 5, where the frozen layer is till the first Max-pooling layer for each input signals. The processing times are 830.15 seconds, 835.64 seconds, and 850.44 seconds, respectively for operation conditions 1 to 3. The Baseline Model showed a processing time of 848.73 seconds. 
These findings indicate the variability in processing time across different models, and emphasis on the effectiveness of the TL based models for efficient training. 

\begin{table}[h!]
\centering
\caption{Total Trainable Parameters and Computational Time for the training phase of the Baseline Model and TL Models.}
\label{tab:trainable_params_computation_time}
\small
\begin{tabular}{@{}llcc@{}}
\toprule
\textbf{Condition} & \textbf{Model} & \textbf{Trainable Parameters} & \textbf{Process Time (s)} \\ 
\midrule
Baseline Condition & Baseline Model & 47,981,011 & 848.73 \\ 
\midrule
\multirow{3}{*}{Condition 1} & Model 1 & 47,962,003 & 185.09 \\ 
 & Model 2 & 47,981,011 & 830.15 \\ 
 & Model 3 & 47,962,003 & 183.03 \\ 
\midrule
\multirow{3}{*}{Condition 2} & Model 1 & 47,962,003 & 184.01 \\ 
 & Model 2 & 47,981,011 & 835.64 \\ 
 & Model 3 & 47,962,003 & 182.94 \\ 
\midrule
\multirow{3}{*}{Condition 3} & Model 1 & 47,962,003 & 185.59 \\ 
 & Model 2 & 47,981,011 & 850.44 \\ 
 & Model 3 & 47,962,003 & 184.13 \\ 
\bottomrule
\end{tabular}
\end{table}
\section{Discussion}
This study proposed a multimodal approach for bearing fault classification that combines both motor phase current and vibration signals within a 1D CNN architecture. The model leverages a late fusion method to fuse features from both modalities. Under the baseline condition (1,500 rpm, 0.7 Nm load torque, and 1,000 N radial force), the model reached an accuracy of 94\% without L2 regularization. After introducing L2 regularization, accuracy rose to 96\%, demonstrating model’s greater resilience against overfitting.

A significant finding of this study is that combining vibration and motor phase current data outperforms previous single-modal approaches. Methods limited to vibration or current data alone often fails to detect subtle fault indicators. By uniting both signals, the model uncovers deeper fault features more effectively, leading to improved fault distinguishability. This multimodal approach improves classification accuracy by several percentage points, reinforcing the benefits of utilizing multiple data sources for fault detection

Another critical finding of this research involves the transfer of previously acquired knowledge from the baseline domain condition to three distinct operating conditions. The key parameters from the baseline condition were altered in the other conditions: a reduction of rotational speed from 1,500 rpm to 900 rpm (a 40\% decrease), load torque dropped from 0.7 Nm to 0.1 Nm (an 85.7\% reduction), and a cut in radial force from 1,000 N to 400 N (60\% less). Conventional fault diagnosis models often struggle with major parameter shifts. In contrast, the pre-trained model effectively adapted to variable operating conditions, demonstrating superior robustness. Among the tested strategies, Model 2, which preserves parameters up to the first max-pool layer and then recalibrates subsequent layers, achieves the best results. Its accuracy remains strong across these new conditions, surpassing the results of directly applying the baseline model to unseen cases. This outcome shows that the model’s extracted features at lower layers, initially tuned at 1,500 rpm and 0.7 Nm load torque, still prove valuable after dramatic parameter changes.

To evaluate our model's performance, we compared the results of our proposed model with existing work, as shown in Table~\ref{tab:methods_accuracy}. From the table, it can be observed that the Central Moment Discrepancy (CMD) method [26] achieved high accuracy (e.g., 94.43\% for Baseline $\to$ Operating Condition 2) by employing domain adaptation to align feature distributions across operating conditions. However, CMD relies solely on vibration signals, which limits its ability to leverage complementary features—a limitation addressed by our multimodal approach.

Our proposed Model 2 demonstrated superior performance across different operating conditions, achieving accuracies consistently above 90\%. For Baseline $\to$ Operating Condition 2, our model outperformed the best existing reference model (CMD) by achieving an accuracy of 96\%, as highlighted in Table~\ref{tab:methods_accuracy}. This improvement demonstrates the effectiveness of integrating complementary modalities in our approach, enabling more robust generalization and enhanced fault classification under diverse operating conditions. 
\begin{table}[h!]
\centering
\caption{Comparison of our best-performing models with existing models from the literature under different operating conditions.}
\label{tab:methods_accuracy}
\resizebox{\textwidth}{!}{%
\begin{tabular}{lllc}
\hline
\textbf{Reference} & \textbf{Methods} & \textbf{Tasks} & \textbf{Accuracy (\%)} \\ \hline
\multirow{6}{*}{\cite{li2021central}} & \begin{tabular}[c]{@{}l@{}}Central Moment \\ Discrepancy (CMD)\end{tabular} & Baseline $\to$ Operating Condition 1 & 70.93 \\
 &  & Baseline $\to$ Operating Condition 2 & 94.99 \\
 &  & Baseline $\to$ Operating Condition 3 & 88.87 \\
 & CNN & Baseline $\to$ Operating Condition 1 & 43.52 \\
 &  & Baseline $\to$ Operating Condition 2 & 94.11 \\
 &  & Baseline $\to$ Operating Condition 3 & 47.43 \\ \hline
\cite{gao2023fault} & \begin{tabular}[c]{@{}l@{}}Attention Adversarial \\ Transfer Networks (AATN)\end{tabular} & Baseline $\to$ Operating Condition 1 & 83.00 \\ \hline
\multirow{6}{*}{\cite{huo2023class}} & 1D-CNN & Baseline $\to$ Operating Condition 1 & 50.08 \\
 &  & Baseline $\to$ Operating Condition 2 & 85.08 \\
 &  & Baseline $\to$ Operating Condition 3 & 50.29 \\
 & \begin{tabular}[c]{@{}l@{}}Maximum Classifier \\ Discrepancy (MCD)\end{tabular} & Baseline $\to$ Operating Condition 1 & 65.81 \\
 &  & Baseline $\to$ Operating Condition 2 & 94.43 \\
 &  & Baseline $\to$ Operating Condition 3 & 69.34 \\ \hline
\multirow{3}{*}{\textbf{Our Proposed Model}} & \multirow{3}{*}{\textbf{Model 2}} & Baseline $\to$ Operating Condition 1 & 92.00 \\
 &  & Baseline $\to$ Operating Condition 2 & \textbf{\textcolor{red}{96.00}} \\
 &  & Baseline $\to$ Operating Condition 2 & 89.00 \\ \hline
\end{tabular}%
}
\end{table}

However, an examination of different knowledge transfer approaches reveals uneven effectiveness. Model 2’s selective parameter adjustments after the initial layers attain the best balance between knowledge retention and adaptation. More conservative methods freeze nearly all layers, limiting adaptability, while more aggressive methods leave most parameters unfrozen, risking the loss of previously learned features. This observation highlights the importance of identifying an optimal parameter-freeze strategy.

A trade-off arises between computational cost and performance. Although Model 2 attains high accuracy under all tested conditions, it requires more time to train due to its larger number of parameters. When computational resources are scarce or conditions require a rapid shift, a less resource-intensive approach may offer a practical compromise. For example, freezing additional layers shortens training time but may lower accuracy by a few percentage points.

From an application perspective, these findings hold promise for industrial maintenance and oversight. The model reaches about 96\% accuracy under the baseline condition and adapts well to major parameter shifts without expensive retraining. This improvement cuts downtime and costs, and it eases the deployment of predictive maintenance solutions across diverse and frequently altered industrial environments.

\subsection{Future work}
Our study contributed significant insights into bearing fault classification using novel multimodal 1D CNN model with late fusion and transfer learning-based models. However, it is imperative to acknowledge certain challenges, as these will be cornerstone for future research:

\begin{itemize}
    \item Model explanation methods: Adapting explainable AI methods, such as Model-Agnostic Explanations (LIME), Shapley Additive Explanations (SHAP), and Saliency Maps could aid in understanding the intricate details of the complex behavior of CNN based models. Such investigation could enhance the transparency and interpretability of the complex models, thus, making decision-making processes more reliable and trustworthy in bearing fault diagnosis.
    \item Exploring diverse architectures: In our study, we have used three TL based models by fine-tuning and freezing layers on the proposed Baseline Model. Employing attention mechanisms within CNN might enable the model to focus on the most relevant part of the input signals and assist the model to precisely identify bearing faults. 
    \item Addressing Class Imbalance: In order to improve the classification of underrepresented class types, future research could use different data balancing techniques like data augmentation, over- and under-sampling, and leveraging class weights. 
    \item Diverse Datasets: The proposed methodology could be evaluated on other publicly available bearing fault datasets. This could be useful in validating the model’s adaptability and generalization capabilities across varied datasets. 
    
\end{itemize}

\section{Conclusion}
This work introduces a multimodal 1D CNN approach that uses vibration and motor phase current signals for bearing fault classification under variable operating conditions. By using late fusion, the model draws on complementary information from multiple sources. After applying L2 regularization, the model reaches 96\% accuracy under the baseline setting, a clear improvement over the non-regularized version. Employing Transfer Learning (TL) strategies enables the model to maintain strong performance under significantly altered rotational speeds, load torques, and radial forces. Although one TL variant requires more computational effort, it provides the best balance between retaining previously acquired knowledge and adapting to new conditions. From a practical standpoint, these findings point toward a robust, accurate, and adaptable solution that reduces the frequency of retraining, thereby decreasing downtime and lowering operational costs. Future studies may investigate explainable AI methods, new architectural designs, and advanced data balancing techniques to further improve model interpretability, reliability, and applicability across a broader spectrum of scenarios.
\section*{Conflict of interest}
The authors declare no conflict of interest.

\bibliographystyle{unsrt}  
\bibliography{main} 

\appendix
\section{Appendix}
\label{sec:appendix}

\begin{figure}[h!]
    \centering
    \includegraphics[width=0.9\linewidth]{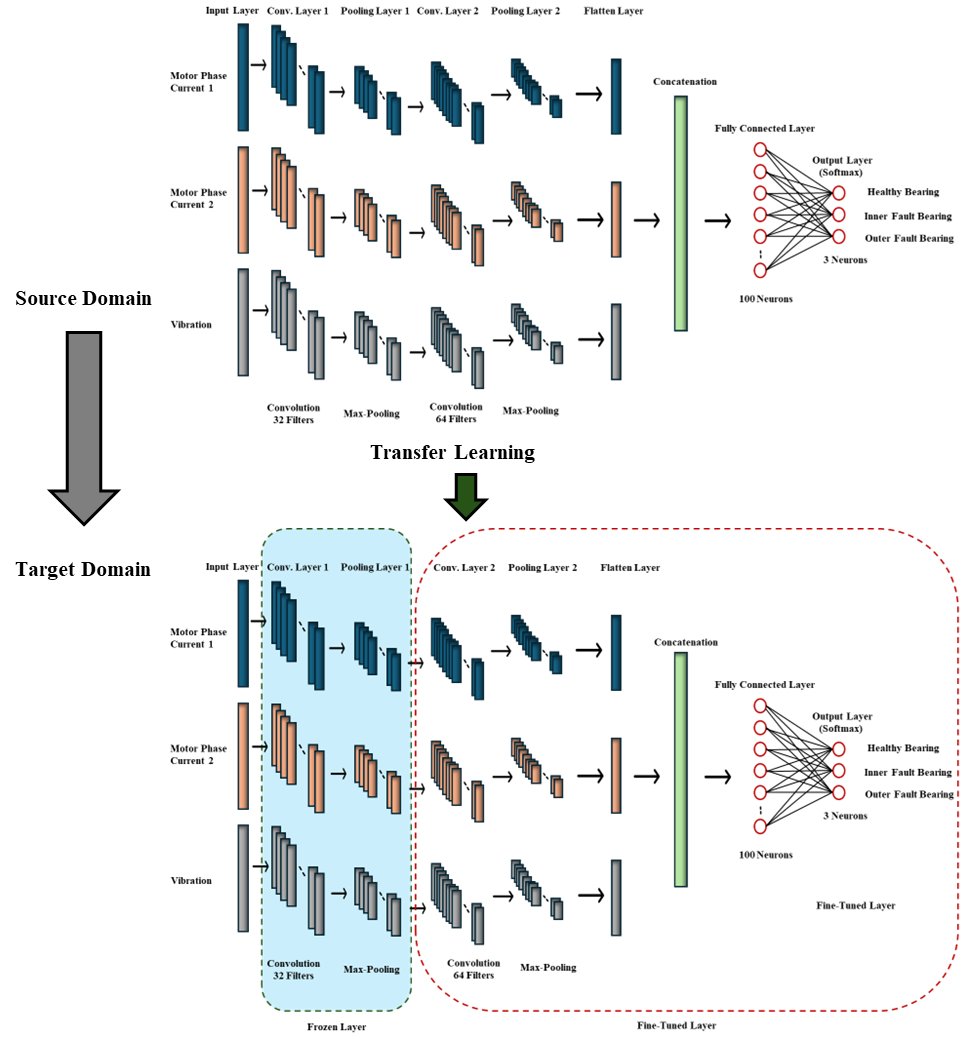} 
    \caption{Framework of proposed Transfer Learning (Model 2)}
    \label{fig:model2}
\end{figure}

\begin{figure}[h!]
    \centering
    \includegraphics[width=0.9\linewidth]{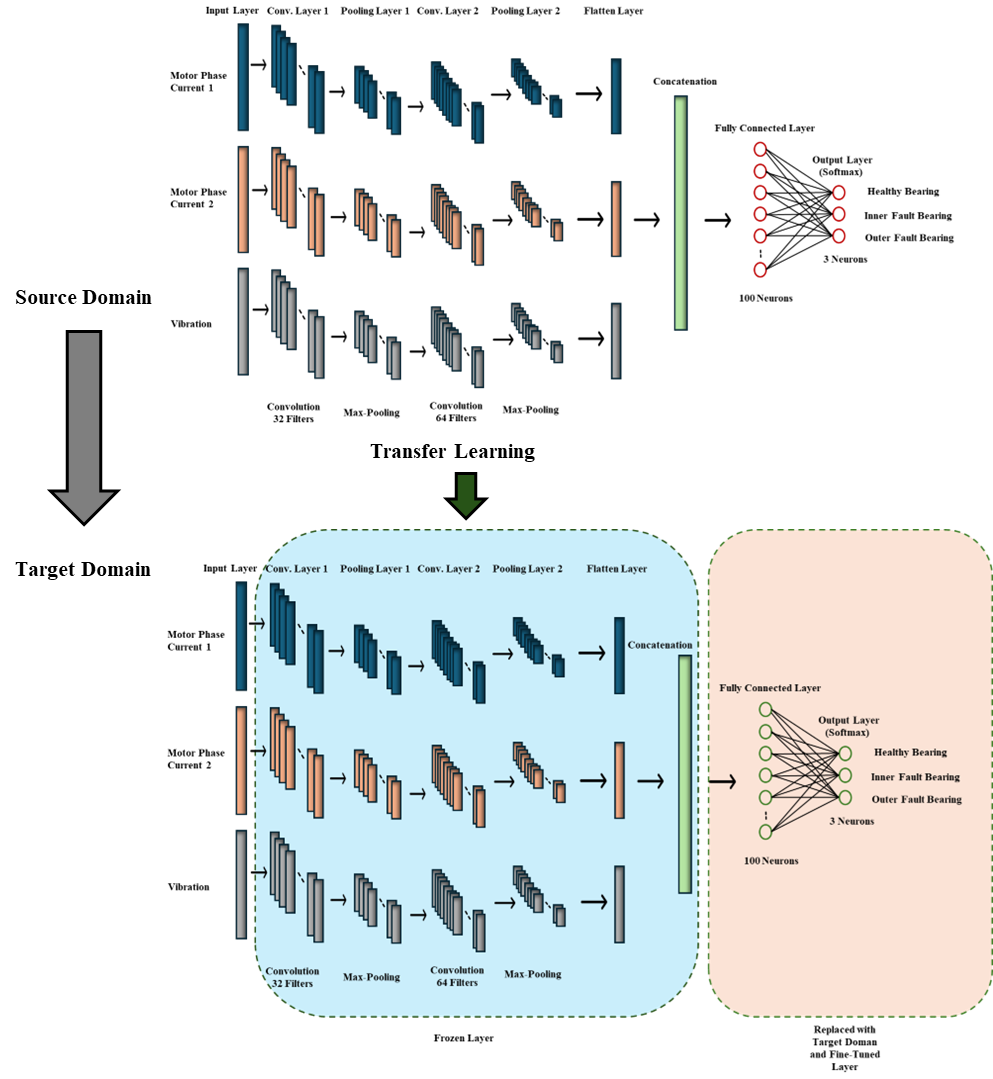} 
    \caption{Framework of proposed Transfer Learning (Model 3)}
    \label{fig:model3}
\end{figure}
\end{document}